\documentclass{osa-article}

\journal{osajournal}

\setprjcopyright

\articletype{Research Article}

\begin{document}

\title{Absolute keV X-ray yield and conversion efficiency in over dense Si petawatt laser plasma}

\author{Sergey~N.~Ryazantsev\authormark{1,2,*}, 
Artem~S.~Martynenko \authormark{2,3}, 
Maksim~V.~Sedov\authormark{2}, 
Igor~Yu.~Skobelev\authormark{1,2}, 
Mikhail~D.~Mishchenko\authormark{2,4}, 
Yaroslav~S.~Lavrinenko\authormark{1,5},
Christopher~D.~Baird\authormark{6}, 
Nicola~Booth\authormark{7},
Phil~Durey\authormark{6}, 
Leonard~N.~K.~Döhl\authormark{6}, 
Damon~Farley\authormark{6}, 
Kathryn~L.~Lancaster\authormark{6}, 
Paul~McKenna\authormark{8}, 
Christopher~D.~Murphy\authormark{6}, 
Tatiana.~A.~Pikuz\authormark{2,9}, 
Christopher~Spindloe\authormark{7}, 
Nigel~Woolsey\authormark{6} 
and Sergey~A.~Pikuz\authormark{1,2}}

\address{\authormark{1}National Research Nuclear University MEPhI (Moscow Engineering Physics Institute), 115409 Moscow, Russia\\
\authormark{2}Joint Institute for High Temperatures of the Russian Academy of Sciences, Moscow 125412, Russia\\
\authormark{3}Plasma Physics Department, GSI Helmholtzzentrum fur Schwerionenforschung, 64291 Darmstadt, Germany\\
\authormark{4}European XFEL, 22869 Schenefeld, Germany\\
\authormark{5}Moscow Institute of Physics and Technology, State University, 141701 Moscow, Russia\\
\authormark{6}York Plasma Institute, Department of Physics, University of York, York YO10 5DD, United Kingdom\\
\authormark{7}Central Laser Facility, STFC Rutherford Appleton Laboratory, Didcot OX11 0QX, United Kingdom\\
\authormark{8}Department of Physics, SUPA, University of Strathclyde, Glasgow G4 0NG, United Kingdom\\
\authormark{9}Open and Transdisciplinary Research Initiative, Osaka University, Osaka 565-0871, Japan
}
\email{\authormark{*}SNRyazantsev@mephi.ru} 



\begin{abstract}
Laser-produced plasmas are bright, short sources of X-rays often used for time-resolved imaging and spectroscopy. Absolute measurement requires accurate knowledge of laser-to-x-ray conversion efficiencies, spectrum, photon yield and angular distribution. Here we report on soft X-ray emission from a thin Si foil irradiated by a sub-PW picosecond laser pulse. These absolute measurements cover a continuous and broad spectral range that extends from 4.75 to 7.5~\AA (1.7--2.6~keV). The X-ray spectrum consists of spectral line transitions from highly charged ions and broadband emission with contributions from recombination, and free-free processes that occur as electrons decelerate in plasma electromagnetic fields. These quantitative measurements are compared to particle-in-cell simulations allowing us to distinguish bremsstrahlung and synchrotron contributions to the free-free emission. We found that experiment and simulation estimations of laser-to-bremsstrahlung conversion efficiency are in a good agreement. This agreement illustrates the accuracy of experiment and physical interpretation of the measurements.
\end{abstract}

\section{Introduction}
Laser produced plasmas (LPP) are widely used as an X-rays source for both fundamental and applied research. This is due to its relatively short duration of the emission and possibility to precisely synchronise a measurement with phase of evolution of the probed object or experiment. It is possible to implement measurement techniques involving X-rays to investigate objects on temporal scales as short as femtoseconds. Broadband emission sources are often required, especially for bioimaging and absorption spectroscopy. LPP at the table-top scale are used as X-ray sources in commercial applications \cite{Kleine2019} and at the very large scale in inertial confinement fusion (ICF) research as very bright X-ray backlighters for capsule explosions. For this purposes petawatt-class (PW) lasers facilities  (such as ARC \cite{Khan2021} and PETAL \cite{Casner2015} or any other mentioned in TABLE E.1 in \cite{OIUL}) able to generate picosecond duration pulses of about kJ are used. Even for ICF experiments, there is a need for bright short-duration sources of soft X-rays. For example, the backlighting of direct-drive cryogenic DT implosions \cite{Sangster2007} uses low energy photons of <2~keV due to low opacity of the plastic shell and deuterium-tritium fuel \cite{Stoeckl2014}. In addition, the low energy part of PW short-duration plasma sources spectrum is used to study warm-dense-matter via absorption spectroscopy \cite{McGuffey2018}.

These two applications, imaging of an ICF experiment and diagnosing warm-dense-matter, requires different spectral composition of the probing radiation. The source with narrow emission band, ideally monochromatic, is required for recording high-quality backlit images. In turn, absorption spectroscopy is most effective when using a source of radiation with continuous spectrum absent of spectral line features or sharp intensity drops \cite{Martynenko2021}. Both types of the probe radiation can be obtained via LPP based due different processes illustrated in Fig.~\ref{figure1}.

Plasma ions produce characteristic spectral lines during transitions of electrons between bound energy levels. For the case of highly charged ions of even relatively low-Z (around 10) the lines lie in keV range of photon energy. Typically, an element is chosen and stripped to the K-shell so that intense emission, usually transitions from the excited state with principal quantum number, n = 2, to the ground state in hydrogen-like (Ly$_\alpha$) and helium-like transition (He$_\alpha$) is at the appropriate wavelength for high-contrast quasi-monochromatic backlighting imaging \cite{Stoeckl2014, Loupias2009}. The plasma spectrum also contains lower intensity continuous emission from free-bound (photorecombination) and free-free (bremsstrahlung and synchrotron emission). The contributions of each process depend on the element and irradiation conditions. As a result, a significant amount of experimental work is needed to characterize a of LPP source, including those produced by picosecond PW pulses. For example, Li et al., \cite{Li2020} estimated photon yield and conversion efficiency from laser- gas jet interaction up to 3$\times$10$^{19}$~W/cm$^2$.

In this paper we investigate the soft X-ray emission properties of near solid density plasma created in a Si foil by a sub-PW laser pulse. The goal of the work is not only to present exact values for the X-ray source characteristics of photon yield in absolute units and angular distribution of the radiation, but also to describe the physical processes contributing to the emission.

\section{Experimental setup}
\begin{figure}[htbp]
\centering\includegraphics[width=13cm]{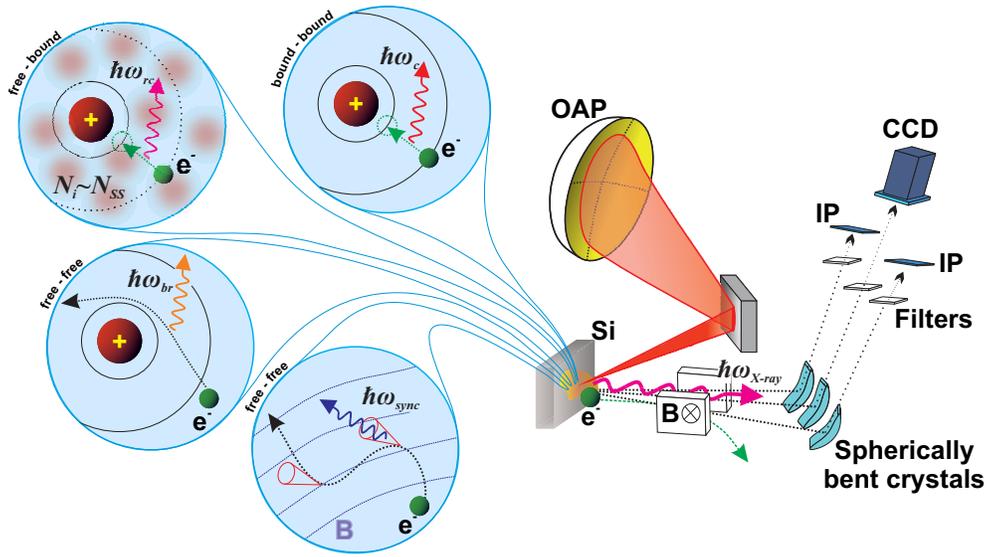}
\caption{Experimental setup and scheme of processes forming the shape of the plasma soft X-ray radiation spectrum registered in the experiments.}
\label{figure1}
\end{figure}

Experimental investigation of emissivity in soft X-ray range as a relativistic-intensity, picosecond duration laser pulse strikes a Si foil was carried out on Vulcan PW laser facility \cite{Danson2004}. A schematic of the experimental setup is shown in Fig.~\ref{figure1}. The plasma was created by irradiation of 2~$\mu$m thick Si foil by laser pulses with intensity of 3$\times$10$^{20}$ W/cm$^2$ ($\lambda$ = 1054~nm, $\tau \approx$ 1~ps, on target energy $\approx$200~J, P = 200~TW) focused on foil surface. The laser beam was focused by an off-axis parabolic (OAP) mirror to a spot with diameter of $\approx$7~$\mu$m with an angle of incidence of 45$^{\circ}$.  The laser contrast was enhanced by using a plasma mirror.

\begin{figure}[htbp]
\begin{minipage}[h]{1\linewidth}
(a)
\center {\includegraphics[width=1\linewidth]{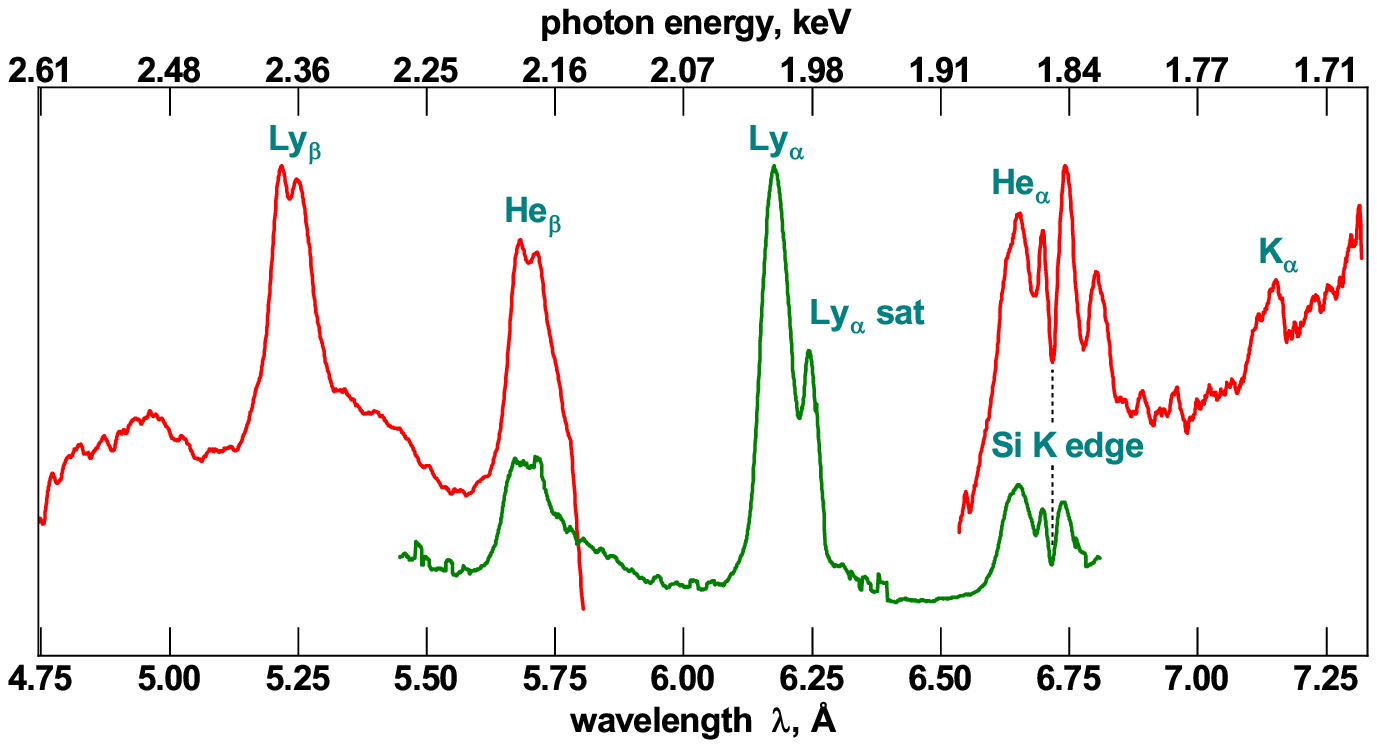}}
\end{minipage}
\vfill
\begin{minipage}[h]{1\linewidth}
(b)
\center{\includegraphics[width=1\linewidth]{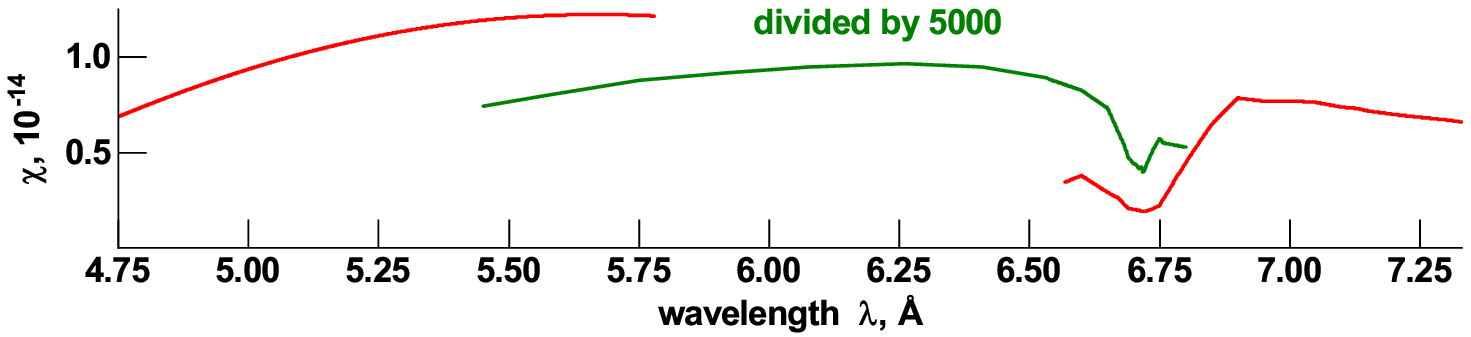}} \\
\end{minipage}
\vfill
\begin{minipage}[h]{1\linewidth}
(c)
\center{\includegraphics[width=0.98\linewidth]{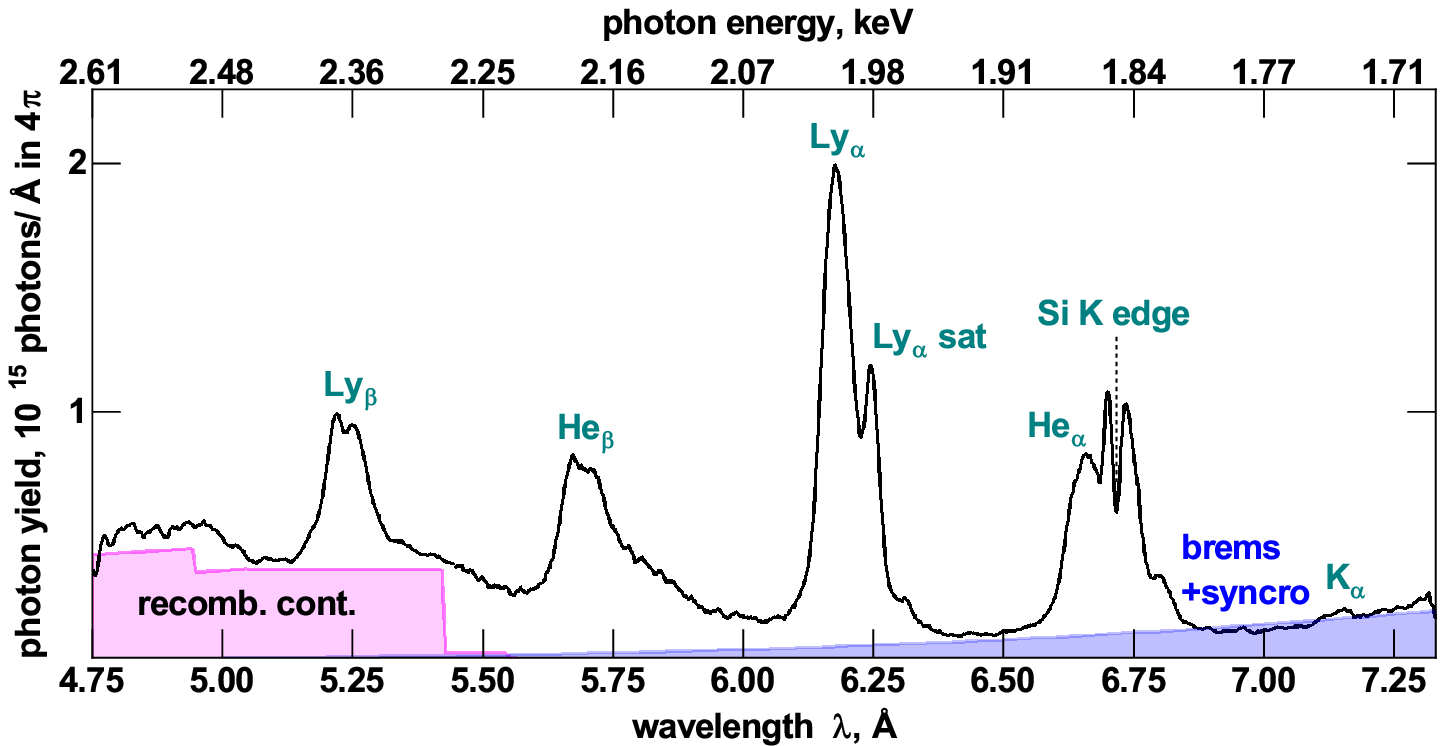}} \\
\end{minipage}
\caption{(a)---normalized experimental emission spectra registered by the FSSRs from the front side of Si foil before convolution with the spectrometer routes response functions $\chi$($\lambda$) shown in (b). $\chi$($\lambda$) is a ratio of the number of registered photons of $\lambda$ wavelength to the one emitted by a source. (c)---combined experimental emission spectrum after convolution by $\chi$($\lambda$). The blue and purple polygons qualitatively demonstrate contributions of bremsstrahlung, synchrotron emission and recombination continuum. Shape of the latter one was calculated for solid-state Si plasma (N$_i$ = 6$\times$10$^{22}$~cm$^{-3}$). }
\label{figure2}
\end{figure}

X-ray emission from a front side of the target was recorded using three focusing spectrometers with spatial resolution (FSSR) \cite{Faenov1994} at angles close to the target surface normal. A magnet was installed between the target and FSSR crystals to help prevent fast electrons produced during irradiation reaching the crystals and causing fluorescence. Fujifilm BAS-TR image plates and an Andor CCD DX-434 (for only one of the spectrometers) were used as X-ray detectors. Their sensitive layers were protected and shielded from visible light by filter foils made of Be, Al and polypropylene of different thicknesses. Observation ranges of the spectrometers overlap to provide cross-calibration between the separate FSSRs enabled measurement of continuous and high-resolution spectra across photon wavelengths from 4.7 to 7.3~\AA (1.7 to 2.6~keV). Raw spectra registered by the individual FSSR spectrometers are shown in Fig.~\ref{figure2}(a).

All the data gathered with the FSSRs were corrected for distance from the LPP, filtering, crystal reflectivity and detector response functions. Numerical modelling \cite{Lavrinenko2015}, via ray tracing through spectrometer, used the actual diffraction profiles or crystallographic rocking curves of the spherically bent crystals installed in the FSSRs. They were calculated with XOP \cite{SanchezdelRio2011} software. Rocking curves for the crystals used in experiments and image plates sensitivity functions are presented in the Supplemental Document uploaded with this Pre-print. Transmission functions calculated using the Henke tables \cite{Henke} for filter foils was also taken into account.  A summary of the response functions for each spectrometer is shown in \ref{figure2}(b). After convolution of the registered spectra with them an initial plasma radiation spectrum was restored (Fig.\ref{figure2}(c)). The most significant corrections were needed around the Si K-edge at 6.7135~\AA \cite{Li1995}. This causes a decrease in reflectivity of the dispersive $\alpha$-quartz (SiO$_2$) crystals in the wavelength region close to the He$_{\alpha}$ line. The detailed He$_{\alpha}$ spectral line structure and the overlap between the two FSSRs in this region enables accurate calibration of the spectrometers. 

\section{Discussion}
\subsection{Registered spectrum description}
The spectrum shown in \ref{figure2}(c) contains both characteristic and continuous emission parts. Its shape is formed by all types of electron transitions: free-free, free-bound, bound-bound (\ref{figure1}). Characteristic emission of the plasma ions is produced by transitions between energy states of H- and He-like Si ions. All of them are broadened due to strong Stark effect. The observed widths correspond to the electron density close to Si solid state value (N$_{SS}\approx$6$\times$10$^{23}$~ cm$^{-3}$). It was achieved both by extremely high value of the laser pulse temporal contrast of 10$^{-10}$, this contrast was improved by a plasma mirror installed in a laser beam path. The most intense lines are the Ly$_\alpha$ (2p$\rightarrow$1s transition in an H-like ion). About 2$\times$10$^{14}$ of photons ($\approx$20\% of the total number) were emitted in the narrow wavelength range from 6.05 to 6.35~\AA~(1.95--2.05~keV), which contains the line itself and its dielectronic satellites. This part of the plasma spectrum is the most suitable for implementation of quasi-monochromatic backlights schemes.

In the He$_\alpha$ region, from 6.53 to 6.89~\AA~(1.8--1.9~keV), the number of photons is smaller by approximately two: $\approx$10$^{14}$. This spectral region contains a broad peak formed by overlapped He$_{\alpha}$ (1s2p $^1$P$_1\rightarrow$1s2 $^1$S$_0$ in He-like ion), intercombination line (1s2p $^3$P$_{2,1}\rightarrow$1s$^2$~$^1$S$_0$) and dielectronic satellites produced by Li-like ions. In Ly$_{\beta}$ (3p$\rightarrow$1s), and He$_{\beta}$ (1s3p~$^1$P$_1\rightarrow$1s$^2$~$^1$S$_0$) lines contain roughly 0.5$\times$10$^{14}$ and 0.7$\times$10$^{14}$ photons respectively. In comparison the K$_{\alpha}$ line at 7.2~\AA~is weak as most of the Si is highly ionised. The mentioned numbers of registered photons correspond to following values of conversion efficiency (CE) of laser energy to X-rays for He$_{\alpha}$, Ly$_{\alpha}$, Ly$_{\beta}$, He$_{\beta}$ lines as $\approx$1.5$\times$10$^{-4}$, $\approx$3.2$\times$10$^{-4}$, $\approx$0.94$\times$10$^{-4}$, $\approx$1.2$\times$10$^{-4}$ respectively.

In the short wavelength region of the spectrum there is a pedestal in intensity, this results from the recombination continuum produced by free-bound transitions of plasma electrons. This spectral region contains unresolved spectral lines corresponding to 4p$\rightarrow$1s transitions in H- and He-like ions. This is due to ionization potential depression \cite{Hoarty2013} in the dense plasma. The shape of the recombination continuum calculated using the Hummer-Mihalas \cite{Hummer1988} model for He- and H-like ions in a solid-density Si plasma is shown in Fig.~\ref{figure2}(c) by the magenta area. Approximately 3$\times$10$^{14}$ recombination continuum photons were registered between 4.75 and 5.4~\AA. Simulations indicate that the recombination continuum extends to 2~\AA. The conversion efficiency to this part of the spectrum is considerable and suggests that the recombination continuum produced by PW solid-density Si plasma is suitable for X-ray absorption spectroscopy in the range $\leq$5~\AA, where the spectral lines are absent.

Other continuous emission results from free-free processes, this contributes to the spectrum through a monotonic growth in intensity in the long wavelength region of the spectrum. This is approximated by an exponential function shown by the blue line in Fig.\ref{figure2}(c). 

\subsection{Modelling of the continuous emission.}
There is no an obvious way to distinguish contributions of bremsstrahlung and synchrotron emission experimentally. It can be done analytically on the base of particle-in-cell (PIC) simulation results. There are several PIC codes that can self-consistently describe both synchrotron radiation and bremsstrahlung  - EPOCH \cite{Arber2015}, OSIRIS \cite{Fonseca2008}, CALDER \cite{Lefebvre2003}, PICLS \cite{Mishra2013} and some others \cite{Wan2017,Wu2018}. These codes were used in a variety of works \cite{Brady2014,Nerush2014,Chang2017,Vyskocil2018,Vyskocil2020} for theoretical investigation of high intensity (from 10$^{19}$ to 10$^{24}$~W/cm$^2$) laser pulse energy conversion into continuous radiation. It was found that synchrotron emission dominates roughly at intensities exceeding $\sim $10$^{22}$~W/cm$^2$, but all these investigations are focused on a short laser pulse (30--120~fs). 

\begin{figure}[htbp]
\centering\includegraphics[]{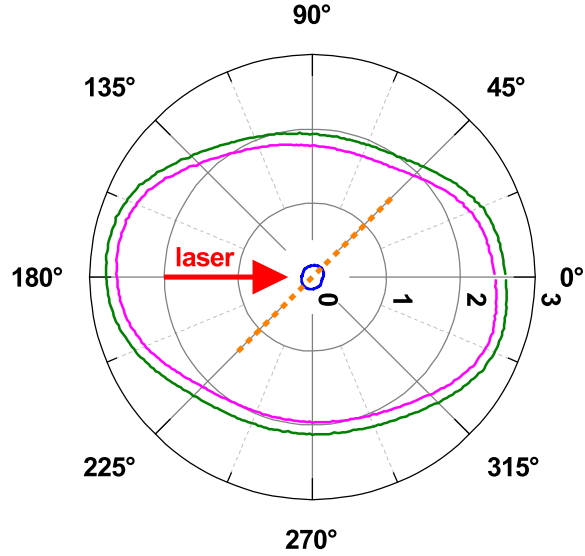}
\caption{Theoretically calculated angular distribution of bremsstrahlung (magenta curve) and synchrotron radiation (blue) emitted in the 4.8--7.3~\AA~(1.7--2.6~keV) wavelength range by a 2 $\mu$m thick layer of fully ionized Si ions with solid-state density irradiated by a subpicosecond laser pulse with intensity 3$\times$10$^{20}$~W/cm$^2$.  The green curve is the sum of two components.}
\label{figure3}
\end{figure}

We have performed two-dimensional (2D) PIC simulation via the EPOCH code for long 700~fs pulse. The code simulates bremsstrahlung radiation using a Monte Carlo method through the elastic and inelastic cross section for the electrons \cite{Vyskocil2018}. Synchrotron emission and radiation reaction are calculated in accordance with principles described in \cite{Duclous2011,Elkina2011}. All the modelling parameters were chosen close to experimental conditions: laser wavelength $\lambda_{laser}$ = 0.8~$\mu$m, focal spot radius 4~$\mu$mm, angle if incidence $\alpha$=45$^{\circ}$, intensity I$_{las}$=3$\times$10$^{20}$~W/cm$^2$. The laser pulse had Gaussian spatial and 3rd order super-Gauss temporal profile. A 2~$\mu$m thick layer of fully ionized Si ions with solid-state density (5$\times$10$^{22}$~ions/cm$^{-3}$) was used as a target. A long laser pulse requires a large simulation box to accurately describe the laser-target interaction and to accommodate the expanding plasma. In our simulations the box size was 120×120~$\mu$m with 10$\times$10~nm grid. The simulation included a current smoothing algorithm and third order particle weighting to limit noise and numerical heating. All boundary conditions were absorbing for radiation and thermalizing for particles. The target was divided into 3 zones: Zone 1 with 50 ion macro-particles per cell (ppci) and 50$\times$14 electron-macro-particles per cell (ppcel), zone 2 with 25 ppci  25$\times$14 ppcel, and zone 3 with 10 ppci , 10$\times$14 ppcel. This allows efficient use of computational time without loss in accuracy. As a result, a temporally integrated angular distribution of bremsstrahlung and synchrotron X-rays in the wavelength range observed in the experiment was obtained, this is shown in Fig.~\ref{figure3}.

In summary the distribution is slightly elongated along the laser beam propagation axis. The ratio of maximum values for back (180$^{\circ}$) and transverse (90$^{\circ}$) directions is $\approx$0.8. For all directions the synchrotron radiation is significantly less intense than the bremsstrahlung. Thus 95\% of $\approx$1.5$\times$10$^{14}$ photons associated with free-free emission corresponds to bremsstrahlung, i.e. integral over the blue curve in Fig.\ref{figure2}(c). It contains about 63~mJ. Therefore, conversion efficiency of laser energy to the bremsstrahlung is $\approx$3.2$\times$10$^{-4}$, which is close to the value $\approx$5$\times$10$^{-4}$ predicted by the PIC simulation.

\section{Conclusions}
The laser plasma produced by sub-PW (1~ps, 200~J, 200~TW, with focal spot diameter 7~$\mu$m, and on target intensity 3$\times$10$^{20}$~W/cm$^2$) laser pulse in a 2~$\mu$m thick Si foil is a very bright source of soft X-rays. We recorded about 10$^{15}$ photons emitted by plasma in the wavelength range of 4.75--7.3~\AA~(1.7--2.6~keV) with total energy $\approx$0.33~J and a CE of about 1.5$\times$10$^{-3}$. About 40\% of this energy were emitted in Si XIV (Si$^{13+}$) Ly$_{\alpha}$, Ly$_{\beta}$ and Si XIII (Si$^{12+}$) He$_{\alpha}$ and He$_{\beta}$ resonance spectral lines and associated satellites. Ly$_{\alpha}$ line is the most intense spectral line containing about a half of all emitted photons. This makes this line the best choice for quasi-monochromatic X-ray backlighter imaging and is sufficiently bright for use in appoint-project Bragg crystal imaging system.

About 170~mJ (CE$\approx$8.4$\times$10$^{-4}$) of incident laser energy was re-emitted in continuous spectrum, of this approximately 64~mJ (CE $\approx$3.2$\times$10$^{-4}$) of it corresponds to bremsstrahlung radiation. The rest is associated with recombination continuum emission. The experimentally obtained laser-to-bremsstrahlung CE is close to 5$\times$10$^{-4}$ predicted by the EPOCH PIC code. The simulations are in qualitative and quantitative agreement with experimental results. The recombination continuum is sufficiently bright and featureless for use as an absorption spectroscopy source.

\begin{backmatter}
\bmsection{Acknowledgments}
The authors would like to thank the Central Laser Facility staff, whose dedication and expertise were essential to the success of their experiment. Calculations were carried out on the computational resources of the JSCC RAS. The reported study was funded by RFBR, project number 19-32-60050. This study was done in the frame of the State Assignment to JIHT RAS (topic \#075-00892-20-00). The work of UK team received financial support from UK EPSRC grants EP/P026796/1, EP/L01663X/1 and EP/H012605/1.

\bmsection{Disclosures}
The authors declare no conflicts of interest.

\end{backmatter}

\bibliography{Ryazantsev_paper_text}

\end{document}


\maketitle

\section{Recalculation of raw data counts to number of real photons}

\begin{figure}[H]
\centering
\fbox{\includegraphics[width=\textwidth]{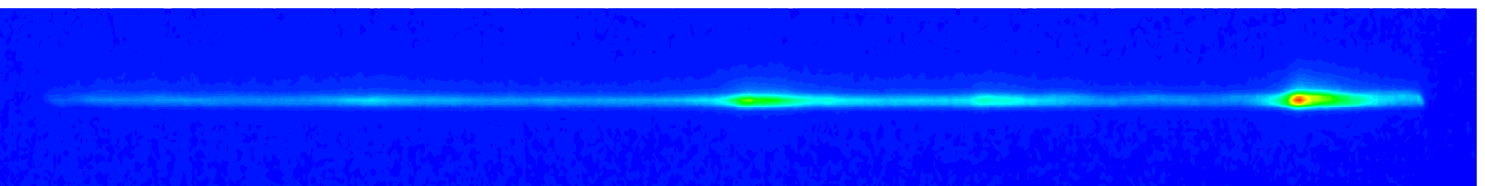}}
\caption{An example of an experimental X-ray emission spectrum registered by the IP detector installed in the FSSR.}
\label{figureA0}
\end{figure}

Raw data recorded by the Fuji BAS TR Image Plates (IPs below) detectors in the experiments is a two-dimensional matrix of numbers. Each of them is proportional to a dose absorbed by a certain pixel of the detector. This matrix can be represented in a form of a grayscale image shown in Fig.~\ref{figureA0}. The spectrum is a profile along the bright narrow field in the centre of the image. This zone has a non-zero width in pixels, so each dot of the spectrum is a sum along the vertical axis.

\begin{figure}[h]
\centering
\fbox{\includegraphics[width=7cm]{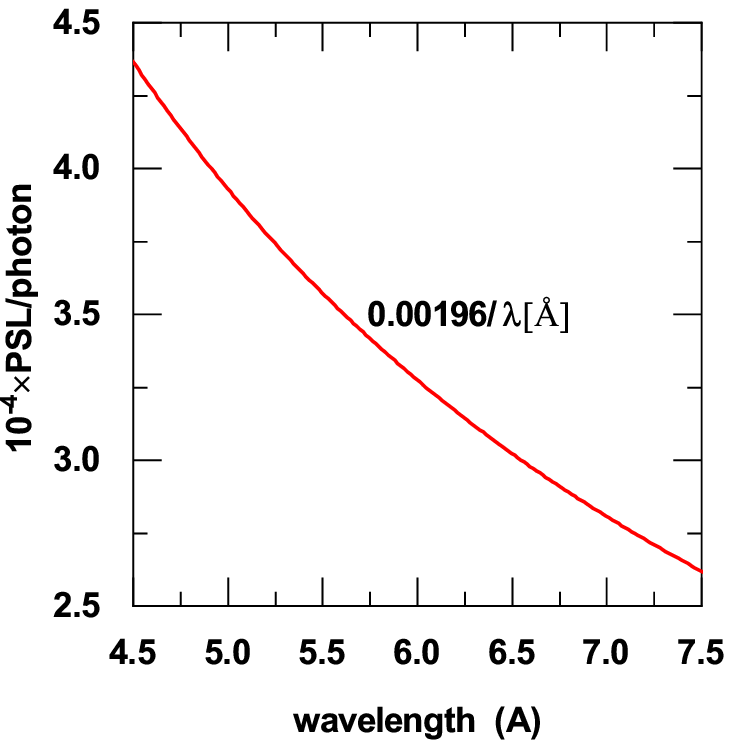}}
\caption{Sensitivity of Fuji BAS TR IPs in the soft X-ray range from \cite{Haugh2013} fitted by a hyperbola function.}
\label{figureA1}
\end{figure}

The data from IPs were digitized via Fujifilm FLA 5100 scanner. A raw image produced as a result of scanning is scaled in arbitrary units (counts), which are related with absolute units called PSL (Photostimulated Luminescence) by the equation \ref{eqS1} \cite{Haugh2013}:

\begin{equation}
PSL = \left(\frac{R}{100} \right)^2\left( {\frac{400}{S}} \right)10^{L\left( {\frac{G}{2^B-1}} - \frac{1}{2}\right)},
\label{eqS1}
\end{equation}
where $R$ is the resolution in microns, $S$ is a sensitivity setting, $L$ is latitude, $B$ is a dynamic range in bits, $G$ is the raw image grayscale value, which is sometimes mentioned as a quantum yield. We used the following values for scanning: $R$ = 25, $S$ = 5000, $L$ = 5, $B$ = 16. It should be noted that the counts-to-PSL recalculation function can be significantly different for other scanner models.  For example, for  General Electric Typhoon FLA 7000 the equation can be found in \cite{Golovin2021}. The values in PSL were recalculated to a number of photons with a given wavelength on the base of a calibration curve shown in Fig.~\ref{figureA1}. 

\section{Crystal reflectivity}
\begin{figure}[H]
\begin{minipage}[h]{0.47\linewidth}
\center{\includegraphics[width=1\linewidth]{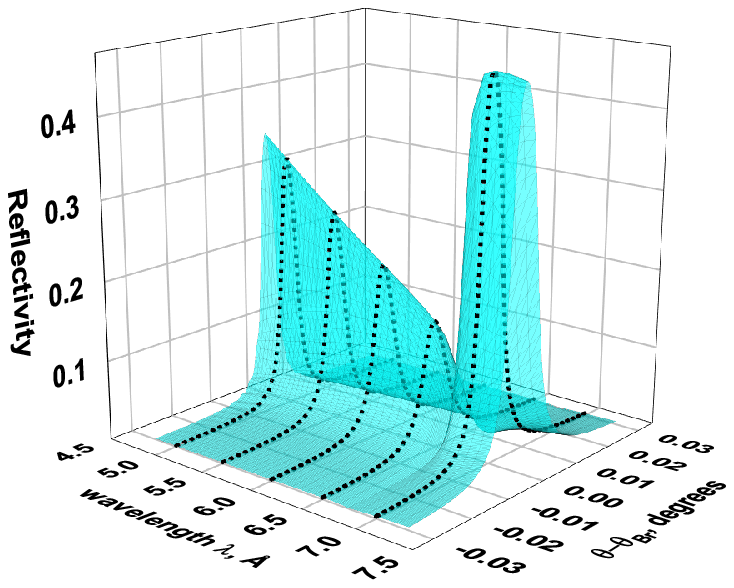}} \textbf{(a)} \\
\end{minipage}
\hfill
\begin{minipage}[h]{0.47\linewidth}
\center{\includegraphics[width=1\linewidth]{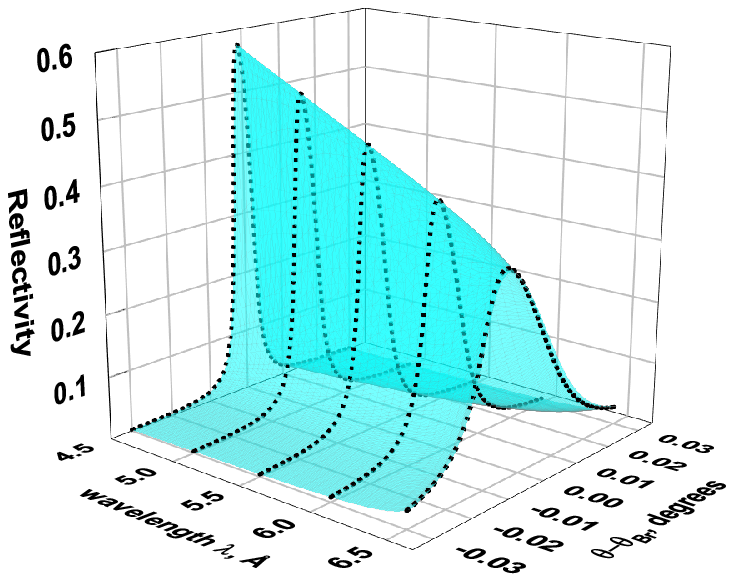}} \\\textbf{(b)}
\end{minipage}
\vfill
\begin{minipage}[h]{0.47\linewidth}
\center{\includegraphics[width=1\linewidth]{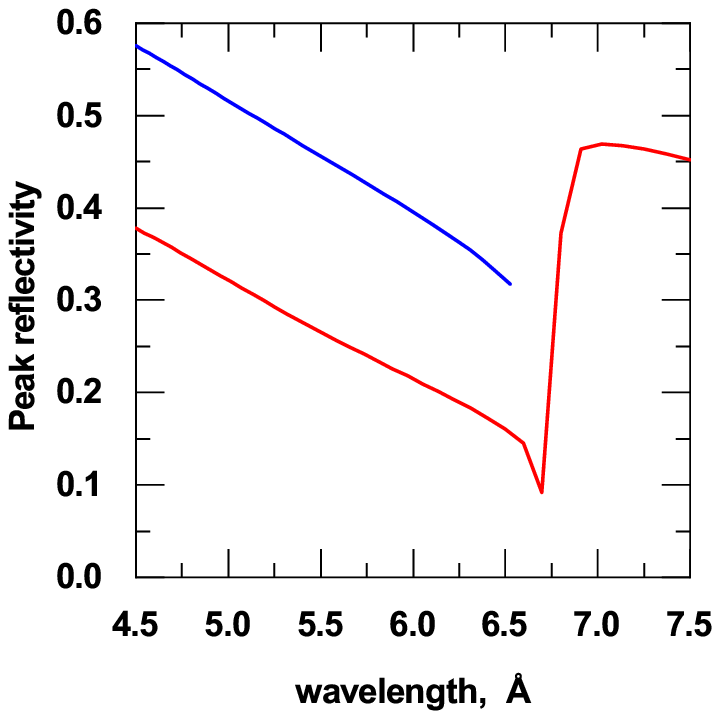}} \textbf{(c)} \\
\end{minipage}
\hfill
\begin{minipage}[h]{0.47\linewidth}
\center{\includegraphics[width=1\linewidth]{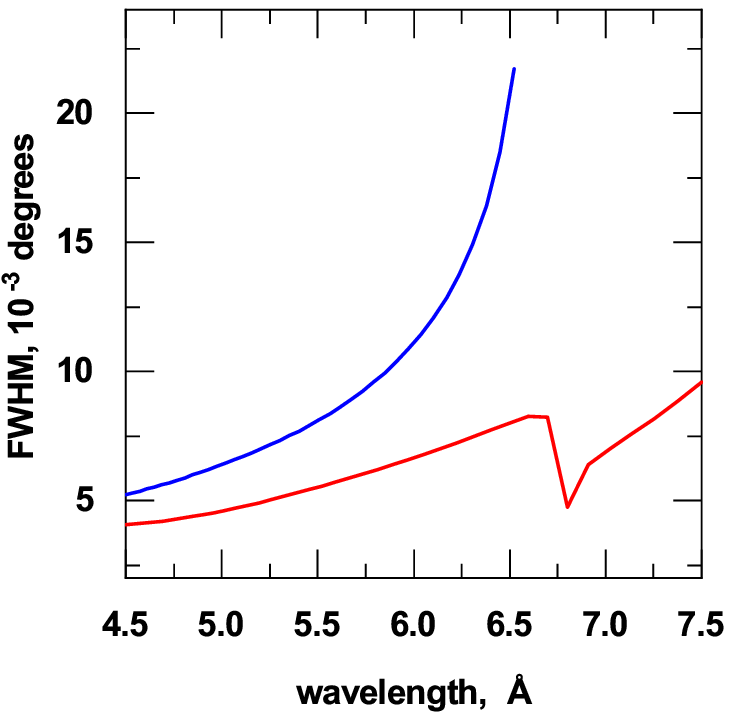}} \textbf{(d)} \\
\end{minipage}
\caption{Set of rocking curves presented as a 3D surface calculated by the XOP software  for spherically bent $\alpha$-quartz crystals with Miller indexes \textbf{(a)} (100) and \textbf{(b)} (101) for the range of wavelengths observed in the experiments. For (101) the data is not presented for $\lambda>6.666$~\AA, because the crystal is not able to reflect photons with a wavelength longer than its interplanar spacing 2d = 6.666~\AA. Dependence of the peaks amplitude (peak reflectivity for a particular wavelength) and FWHM are given on the planes \textbf{(c)} and \textbf{(d)} correspondingly: red line for (100), blue line for (101).}
\label{figure2}
\end{figure}

The corrections associated with Crystal Reflectivity (CR) were also considered during spectra restoring process. The CR function here is N$_r$/N$_e$, where N$_r$ and $N_e$ are numbers of photons reflected by a crystal and emitted by a source, correspondingly. It depends on a solid angle covered by the spectrometer crystal ($\Omega_{cr}$) and also on its rocking curves (diffraction profiles). $\Omega_{cr}$ was calculated directly from the distance between the source and the crystals and the surface dimensions of the latter ones. Rocking curves were obtained by the X-ray Oriented Programs (XOP) \cite{SanchezdelRio2011}. This software allowed to simulate diffraction properties of spherically-bent crystals on the base of extended dynamical theory, which is fully described, for example, in \cite{Authier2003}. The calculated rocking curves are shown in Fig.~\ref{figure2}. The curves were taken into account in a numerical simulation (modelling principles are described in \cite{Lavrinenko2015}) of rays propagation through the spectrometers optical schemes. They were considered as profiles of probability for a photon with given energy and incidence angle to be reflected by spherically bent crystals of the FSSRs.

\bibliography{Ryazantsev_PS_abs_val_appendix}